\documentclass[letterpaper, 10 pt, conference]{ieeeconf}  
\IEEEoverridecommandlockouts                              
\makeatletter
\let\NAT@parse\undefined
\makeatother

\usepackage[dvipsnames]{xcolor}

\newcommand*\linkcolours{ForestGreen}

\usepackage{multirow}
\usepackage{listings}
\usepackage{multicol}
\usepackage{lipsum}

\usepackage{url,hyperref}
\hypersetup{
colorlinks,
linkcolor=\linkcolours,
citecolor=\linkcolours,
filecolor=\linkcolours,
urlcolor=\linkcolours}

\usepackage[labelfont={bf},font=small]{caption}
\usepackage[none]{hyphenat}

\usepackage{mathtools, cuted}

\usepackage[noadjust, nobreak]{cite}

\usepackage{tabularx}
\usepackage{amsmath}

\usepackage{float}

\usepackage{pifont}

\newcolumntype{Y}{>{\centering\arraybackslash}X}

\usepackage[]{placeins}

\usepackage{placeins}

\usepackage{tikz}

\usepackage[framemethod=tikz]{mdframed}

\usepackage{afterpage}

\usepackage{stfloats}

\usepackage{atbegshi}
\newcommand{\handlethispage}{}
\newcommand{\discardpagesfromhere}{\let\handlethispage\AtBeginShipoutDiscard}
\newcommand{\keeppagesfromhere}{\let\handlethispage\relax}
\AtBeginShipout{\handlethispage}

\usepackage{comment}

\title{\textbf{Towards a debuggable kernel design}}
\author{Ashish Gupta, Chandrika Parimoo \{ashishgu,cparimoo\}@andrew.cmu.edu }

\begin{document}

\maketitle
\thispagestyle{empty}
\pagestyle{empty}

\begin{abstract}
This paper describes what it means for a kernel to be debuggable and proposes a kernel design with debuggability in mind. We evaluate the proposed kernel design by comparing the iterations required in cyclic debugging for different classes of bugs in a vanilla monolithic kernel to a variant enhanced with our design rules for debuggability. We discuss the trade offs involved in designing a debuggable kernel.
\end{abstract}

\section{INTRODUCTION}

Operating system kernels generally consist of several subsystems that handle memory management, process management, process scheduling, network, I/O and peripheral devices. The interactions between and within different parts of the kernel are asynchronous. The size of the kernel and the inherent asynchronous concurrent model of programming makes the implementation of a robust kernel challenging. When bugs arise, isolating and reproducing them is generally harder than userspace programs due to the mentioned properties of a kernel. Kernel development and debugging is arcane\cite{DebuggingKernelTutorial}, therefore creating a high barrier of entry for new kernel developers.
Debugging is one of the most costly phases in software development. Developers spend over 30\% of their time debugging and validating software\cite{DebuggingMindset}. The cost of debugging rises up as one moves to lower layers of the software stack because of the increasing amounts of guarantees provided to the layers above. Debugging as an early design consideration can help make the process shorter, alleviating costs.\\

Kernel debugging has conventionally proceeded through the use of tools which stand at various levels of maturity, but never from the perspective of the kernel’s design itself. We define ”debuggability” as the kernel’s inherent ability to lend itself to debugging. We argue that the kernel knows best about itself and can be designed to provide well structured and relevant information about the deviation in its behavior and make the process of debugging the kernel (itself) more efficient. Debuggability can thus be considered an intrinsic property of the kernel and taken into account during the design of the kernel instead of being an afterthought.

\section{RELATED WORK}
Existing debugging approaches include tracing and probing with the use of tools such as DTrace\cite{Dtrace} which allow live patching of instructions with instrumentation code in order to get more information about instructions being executed. This can be a useful technique for debugging, but only captures the execution trace (where overhead can be as large as 100x for high frequency function calls\cite{DtraceOverhead}). Our approach exposes the kernel’s high level information to the developer in the form of data structures that the developer understands instead of only relying on the execution stream and function arguments. \\

Kernel developers use forms of remote debugging (over a serial line)\cite{FreeBSDDoc}. While this technique has been an effective way for kernel debugging, it requires the possession of an additional machine to run a master kernel on one machine sending commands to the debugee kernel on another machine. We do away with the requirement for additional hardware for debugging by incorporating the ability for a kernel to help debug itself. \\

Hardware based approaches such as JTAG\cite{JTag} are very powerful expose a lot of detail about the execution with the help of a special controller. Although one can get a large amount of information , the information provided can be too low-level to be useful for debugging kernel design issues. Our approach, on the other hand, focuses on the kernel’s high level structures providing means to infer some low-level information as well. \\

Hypervisor based record and replay approaches such as Samsara\cite{Samsara} offer an excellent way to tackle hard to reproduce bugs through the use of virtual machines and hypervisors. Although effective, they inherently lack performance as the use of virtual machines and hypervisors slows down performance (upto 6x on a 4-core machine) limiting their usefulness. In our approach, we wish to get rid of the overhead imposed by the use of virtual machines and exploit the hardware the kernel runs on directly to provide a record-and-replay feature as a part of d-mode. \\

Diagnosys\cite{Diagnosis} allows the automatic generation of debugging interfaces by statically analyzing the kernel code for safety holes. This allows Diagnosys to generate relevant logs at run time which help in debugging. The approach is restricted to the identified safety holes and Diagnosys is added as a module to the Linux kernel rather than being integrated in the design of the kernel. \\

The work on exposing bugs before having to debug them provides good examples of the kind of bugs most likely to be found in a kernel. Landslide\cite{Landslide}, in its approach of systematic exploration of concurrency bugs provides insights into which bugs frequently occur in the kernels created at CMU which follow the Pebbles kernel specification. Since we modify Pebbles\cite{Pebbles} to prototype our kernel design, we found Landslide to be useful in guiding our attempts to craft scenarios to reason about the design rules of a debuggable kernel. \\

Our approach differs from conventional tools in the sense that we want to make the kernel a more active participant in the debugging activities of a developer as the kernel has access to all of the information within itself and can be presented in the right way to a developer which can be used to create a mental model of the cause of the issue without the need to actually analyse memory and step-wise debug.\\

\section{DEBUGGABLE KERNEL}

\subsection{Debuggability}

\begin{figure*}
\vspace{2mm}
\centering
\includegraphics[width=0.8\linewidth]{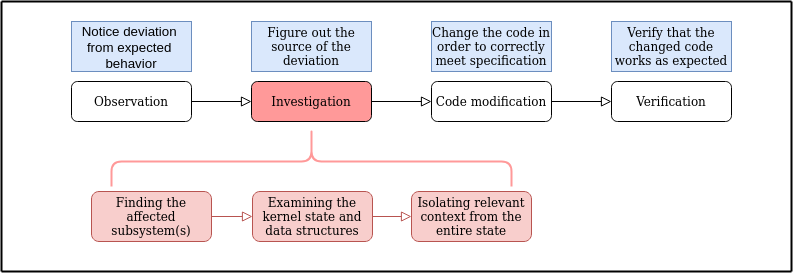}
\setlength{\belowcaptionskip}{-15pt}
\caption{\textbf{Phases of debugging}}
\label{fig:debug_phase}
\end{figure*}

Cyclic debugging is the iterative process of narrowing down the reason for program failure. In each iteration, the programmer gathers more information, adding to the knowledge gained from previous debug iterations, finally zeroing in on the root cause of the bug\cite{DrDebug}. Fig~\ref{fig:debug_phase} outlines the phases typically involved in the process of isolating and fixing a bug. The investigation phase is where most of the developer effort and time is spent. Within the kernel, the investigation phase involves isolating the subsystem (the source of the bug) and isolating relevant context about the subsystem in order to find the root cause of the bug. This typically involves several iterations in cyclic debugging as different aspects of the behavior must be logged and analyzed before branching to investigating any callee functions. The process increases in time and complexity when the bug’s source is the non-deterministic execution within the kernel as it becomes much harder to reproduce, and therefore isolate.\\

We define the debuggability of the kernel for a specific bug as the number of iterations it took to find the root cause of the bug using cyclic debugging. As each iteration of the debugging process involves recompiling and re-executing the kernel, a higher number of iterations makes the ordeal of debugging more expensive thus decreasing the debuggability of a kernel. 

\subsection{Debuggability Scenarios} 

To motivate the design for our kernel, we came up with some common bugs that can creep up during kernel development. These bugs are representative of some of the classes of bugs encountered during kernel development. This allows us to focus on one class of bugs at a time, and suggest features in a kernel’s design to combat those issues. We use the experience gained through the process of debugging these issues to come up with ”debuggability rules”. We claim that a kernel designed with these rules in mind is more debuggable. \\

\subsubsection{\underline{Memory Exhaustion}}

In Listing 1, child processes are forked and the expectation from the program is to print the message ”Another one bites the dust” endlessly, potentially with a few ”Fork failed” messages. We introduce a bug in the kernel’s wait() syscall implementation by removing free for the PCB of the waited for process leading to a memory leak. This results in memory exhaustion which is evident from the program’s behavior: It initially prints ”Another one bites the dust” several times, but soon after only prints ”Fork failed”. While debugging we felt the need for a global view of kernel data structures in memory, access to which would have led us straight to the process subsystem. Since wait() is responsible for cleaning up the exited children, there must be an issue in the implementation of wait(). This led us to the first rule. \\

\textbf{Debuggability Rule 1}: A kernel which provides visibility into its data structures and resource usage improves its debuggability. Ability to provide a debugging interface under memory exhaustion increases debuggability as otherwise it is hard to debug the kernel in such situations.

\begin{lstlisting}[language=C, caption=Fork bomb, flexiblecolumns=true, fontadjust=true, frame=single]
// wait endlessly to reap exiting children
void reaper(void) {
    while(1) {
        wait(NULL);
    }
}

// fork endlessly
void bomberman(void) {
    while(1) {
        pid_t pid = fork();
        if(pid == 0) {
            printf(``Another one bites the dust\n'');
            exit(0);
        } else if(pid == -1) {
            printf(``Fork failed\n'');
        }
    }
}
// Spawn two threads, one doing all the fork()ing and
// the other doing all the wait()ing
int main() {
    int tid = thr_fork();
    if(tid == 0) {
        bomberman();
    }
    else {
        reaper();
    }
}
\end{lstlisting}

\hfill\\
\subsubsection{\underline{Use After Free}}
In Listing 2, a thread maps pages into its process address space and tries to make the kernel copy file data into these pages using the readfile syscall. Another thread unmaps the same pages in a loop to occasionally cause readfile to fail and print the message ”Failed to read file” and occasionally ”Someone’s acing the exam”. The kernel should ideally check the virtual memory mappings of user supplied pages in its readfile() implementation. We introduce a bug by removing any such validation and one of the inter leavings on the run results in readfile() trying to write file data into an unmapped memory address causing the kernel to page fault, resulting in a kernel panic. In an attempt to debug this, we enabled the logging of all VM operations which would help us track this issue down. Once we had the logs, it was difficult to make sense of them and we had to use external tools to process the logs into a readable format, keeping the value of cr2 handy. We wondered if the kernel itself could carry out all these steps, freeing us of the overhead of making sense of this data. Based on this experience, we invented another rule. \\

\textbf{Debuggability Rule 2}: Keep history of the actions performed by a subsystem along with the identifying information about the requesting process / thread. This history should be available for navigation by a user in the event of a kernel panic.

\begin{lstlisting}[language=C, caption=Pulling the rug from under the kernel's feet, flexiblecolumns=true, fontadjust=true, frame=single]
\label{lst:UAF}
void remove(void *addr) {
    // continuously remove pages to 
    // create many possible inter-leavings
    while(1) {
        remove_pages(addr);
    }
}

int main() {
    void *addr = 0x13370000;
    int size = 4096;
    pid_t tid = thread_fork();

    // newly spawned thread should do 
    // remove_pages endlessly
    if(tid == 0) {
        remove();    
    }

    while(1) {
        int failed = new_pages(addr, size);
        if(failed)
            continue;

        // send buffer to kernel to read 
        // in contents of a file
        int code = readfile("exam_solution.txt",
            addr, size, 0);
            
        if(code < 0) {
            printf``Failed to read file\n'');
        }
        else {
            printf(``Someone`s acing the exam!\n'');
        }
    }
\end{lstlisting}

\hfill\\
\subsubsection{\underline{Non determinism}}

Bugs that surface due to asynchronous interactions between kernel components are the hardest to tackle. In Listing 3, a process’ thread invokes the yield() syscall to yield the CPU to a different thread. We analyse an interleaving where the timer fires when the kernel is processing the yield() request: the kernel has obtained the caller’s TCB but the timer interrupt handling results in the other thread of same process being scheduled. The other thread exits deleting its TCB, while the kernel returns execution to the original invocation of the yield syscall and continues the process of switching to the thread would be to find the exact interleaving of events that causes the issue. As logging alone seems ineffective at tackling such bugs, it seemed crucial that there be a mechanism in the kernel which did most of the heavy lifting required for investigation. This led us to the final rule. \\

\textbf{Debuggability Rule 3}: A kernel which allows a replay mechanism which performs operations exactly in the order as they happened in the original run has higher debuggability than a kernel which does not. A mechanism must exist that can reliably reproduce a bug for investigation even if the initial cause of the bug was non deterministic inter leavings in the kernel’s execution.

\begin{lstlisting}[language=C, caption=Dead Yield, flexiblecolumns=true, fontadjust=true, frame=single]
// Code running in userspace
int main {
    int tid = thread_fork();
    
    if(tid == 0) {
        // This is thread 0
        vanish();
    }
    else {
        // This is thread 1
        int failed = yield(pid);
        if(failed)
            printf(``Yielding failed\n'');
    }
}

// Code in kernel space
// Implementation of vanish system call
// Called by thread 0 from above
void vanish_impl(void) {
    pcb_t *my_pcb = find_pcb_by_tid(gettid());
    remove_from_lists(pcb);
    unmap_destroy_process(pcb);
    run_some_other_thread();
}

// Implementation of yield system call
int yield_impl(int tid) {
    pcb_t *target_pcb = find_pcb_by_tid(tid);
    // -----> Timer interrupt fired here
    
    // do the context switch to target_pcb
    stack_switch(target_pcb);
    pop_gp_registers();
    return_to_user_mode();
}
\end{lstlisting}

\subsection{Design and Implementation}
Based on the experiences gathered by analyzing the debugging process of the bugs described in the previous section, we designed an extension ”d-mode” to the Pebbles\cite{Pebbles} kernel, the combination of which we call ”dPebbles”. d-mode lives within the kernel and records various pieces of information during the run of the kernel so that it can present this information later to a developer for debugging. We now describe different components of d-mode which work together to provide the full debugging experience. d-mode currently runs on x86 hardware, and therefore some design is dependent on features of the x86 hardware. These features are commonly found on most architectures and d-mode can be easily ported to other architectures as a result.

\subsubsection{\underline{debugMode}}
When the kernel panics, or a special key combination is pressed, d-mode presents itself interactively to the user in a familiar shell which serves as a single point of control. The shell accepts various commands, and lists out the information the user asks for. A stack switch is performed to d-mode’s own stack to not have stack overflows or overwrite data on another process’ stack.\\

\subsubsection{\underline{Kernel data Structures}}
d-mode augments the allocation routines of a kernel (through the use of a wrapper) and maintains a count for the number of data structure objects allocated. Only the number of allocated objects needs to be tracked as it can be converted to the actual number of bytes in use by multiplying with the size of the relevant object type. This information serves as an overview of the entire system and can help guide development efforts into the right kernel subsystem. More complex views around this theme can be developed for more fine grained view into the kernel’s memory usage. \\

\subsubsection{\underline{Virtual Memory operations}}
Kernel panics often result from accessing unmapped memory regions in a process’ address space while in kernel mode (Page fault in kernel mode). d-mode deals with these by facilitating viewing past interactions between a process and the VM subsystem. Each function of the Pebbles VM subsystem was augmented by the addition of a function call which records the type of VM operation being performed along with the calling process / thread identifiers and the address being modified. This information is maintained within d-mode and can be viewed by the user when the kernel panics due to a page fault. As d-mode is currently based on x86 systems, the cr2 control register (which contains the address whose access resulted in a page fault) is also shown to the user to have higher debuggability. We use the following format for maintaining VM operations:

⟨ pid, tid, VM operation ID, Virtual Address ⟩ \\

\subsubsection{\underline{Safety}}
When a kernel is buggy, all bets are off as it can overwrite critical data structures (for example by writing data to wild pointers). In order to ensure that a user can still access d-mode, it is imperative to have a reserved memory area that remains safe for use even when the kernel can no longer be trusted. The history of VM operations, as well as the counters for data structures remain resident in this safe area. It must also be ensured that d-mode has enough working area apart from the storage space in order to perform its own functions. As a result, d-mode has a reserved space for using as a stack area. The stack pointer (ESP) is switched whenever entries and exits occur to d-mode’s own stack. It must also be possible for a user to quickly identify if certain data structures have been corrupted. d-mode ensures this through the use of contract validators where it can use kernel exported interfaces to ensure (on a best effort basis) that data structures are intact. \\

\subsubsection{\underline{Non Determinism}}
Please note that the current implementation does not contain this record-and-replay module. We still present the design here for completeness. A kernel being highly concurrent through the occurence of interrupts and interactions with the userspace (via syscalls) results in a class of bugs which are a result of complex interleavings of various subsystems which can be highly elusive to track down. In order to deal with such non determinism, d-mode has a record-and-replay module which runs alongside the kernel collecting enough information in order to be able to restore the kernel to an arbitrary past state (”replay”), ensuring that the kernel progresses exactly in the manner it progressed when the elusive bug was encountered. Every time an ISR (interrupt service routine) is executed, the EIP of the interrupted instruction is recorded along with the ECX register (as the interrupted instruction could be within a loop and therefore different instructions could have the had the same EIP but did not get interrupted) is recorded along with any input that the ISR received (such as a key stroke). During replay, we make use of debug registers (on x86) in order to set breakpoints at recorded EIP locations where the kernel’s execution was interrupted. d-mode then modifies the registers to result in a stack switch and restarts execution after setting up the ISR’s stack so that it does exactly what it did previously. On each breakpoint, control is returned to the user for examination so that the user can set additional breakpoints or look into data structures or memory akin to a conventional debugger (like KGDB\cite{kgdb}). \\

\subsubsection{\underline{Interrupts}}
While executing, d-mode disables all interrupts apart from the keyboard interrupts as that is required for user interaction. While replaying it mimcs the interrupt arrival but reconstructing the interrupt stack from the records.

\section{EVALUATION}
We evaluate the debuggability score of d-Pebbles and Pebbles and the memory overhead incurred for maintaining d-mode’s state in d-Pebbles. The Pebbles kernel only allows a single process to perform allocations, deallocations to protect the memory manager’s data structures. We increment, decrement the data structure counters within these locks and observed that the addition of these counters had no execution overhead. As a result, we only present evaluation for the execution overhead of d-Pebbles incurred by the VM logging module using two benchmarks: 712bench and mutex712bench \cite{d-Pebbles}. \\

\subsection{Debuggability}
We measure the number of iterations it takes for d-Pebbles to debug issues in each of the scenarios listed in Debuggability Scenarios. We use these iteration counts in cyclic debugging of the listed bugs to quantify the debuggability of each kernel. Table~\ref{tbl:comparison} shows the resulting Debuggability scores. We find that d-Pebbles improves the debuggability of Pebbles by at least 66.6\%.
\\

\subsubsection{Memory Exhaustion}
\hfill\\
\underline{Pebbles} \\ 
Iteration 1: Observe fork() fails and turn on debugging in fork()\\
Iteration 2: Observe fork failure due to memory exhaustion, turn on logging in VM\\
Iteration 3: Observe no deallocations of PCB, infer that the bug must be in wait() \\
\underline{d-Pebbles} \\
Iteration 1: Get memory stats and running processes after observing that fork() failed due to memory exhaustion. Large number of PCB allocations and no running processes suggests memory leaks in the implementation of wait() \\

\subsubsection{Use after free}
\hfill\\
\underline{Pebbles} \\
Iteration 1: Examine the crash dump from kernel panic and observe that readfile() caused the page fault \\
Iteration 2: Enable syscall and VM logging to relate the faulting buffer’s address and what operations were conducted on that address \\
Iteration 3: Re-run the kernel and examine the logs to find the use-after-free issue \\
\underline{d-Pebbles} \\
Iteration 1: Drop into d-mode after the kernel panics. Check the VM operations on the address in the cr2 register, EIP from readfile syscall. Note that the faulting address was deallocated by another thread from the VM operation logs, infer that readline does not handle use-after-free issues \\

\subsubsection{Non-determinism}
\hfill\\
\underline{Pebbles} \\
Iteration 1: Examine stack trace of the crashing process and enable logging in the stack switch code path along with yield. \\
Iteration 2: Wonder what might have caused the error, and re-run the kernel with logging enabled.\\
Iteration 3: ... Unable to reproduce the exact bug, enable more logging in the system to understand what’s going on and run the system with logging enabled to catch the bug when it resurfaces. \\
Iteration X: Observe that the timer was fired in between the yield implementation and vanish was called shortly afterwards. Fix the issue. \\
\underline{d-Pebbles} \\
Iteration 1: Drop into d-Mode as a result of a kernel panic. Observe that the crash happened due to yield() and stack switch. \\
Iteration 2: Go back far enough to a point in the past (N time earlier) and rerun the kernel adding breakpoints to understand what might have caused the bug. Repeat with shorter time intervals to the past. \\
Iteration log(N): After binary searching over the time space of the past N time units, should be able to zero in on the bug and fix the issue. 

\begin{table}
\begin{center}
\begin{tabular}{ |c|c|c|c| } 
\hline
Bug Class & d-Pebbles & Pebbles \\
\hline
Memory Exhaustion & 1 & 3 \\ 
Use After Free & 1 & atleast 3 \\ 
Non-Determinism & log(N) & Unbounded \\ 
\hline
\end{tabular}
\caption{\label{tbl:comparison}Debuggability Score.}
\end{center}
\end{table}

\subsection{Memory Overhead}
d-Pebbles reserves an additional 20.216 KB for supporting d-mode. This 1\% overhead comes from a 4 KB kernel stack. for d-Mode, and a 120 bytes array of counters for counting allocations of kernel data structures. The rest is used for a ring buffer for logging VM operations, which we currently limit to a size of 1000 log entries. Although this overhead is fairly small, we expect it to grow when the record and replay engine is implemented in the kernel which would log more operations than just VM operations. The size of the required log will grow to accomodate a larger log of the past operations. \\

\subsection{Execution overhead without contention}
The 712bench performs N VM allocations and N deallocations in a single unithreaded process. We measure the time to perform these N allocations and N deallocations in d-Pebbles with VM logging and compare it with vanilla Pebbles. We use the getticks() system call to measure elapsed time in CPU timer ticks. Fig 2a shows the relative overhead of d-Pebbles by varying the total operations, 2N. We observe a low overhead of less than 1.5\% of d-Pebbles over Pebbles.\\

\includegraphics[scale=0.4]{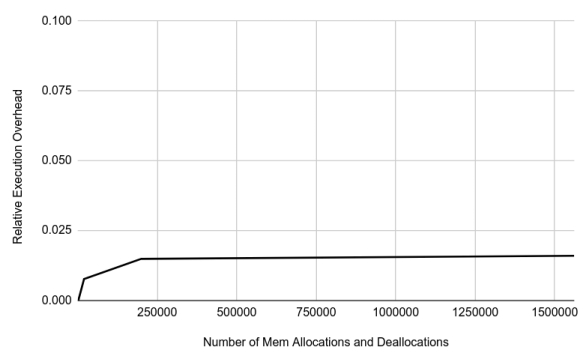}
\begin{center}
    Fig 2a. Execution Overhead of d-Pebbles for 712bench
\end{center}

\hfill\\
While running the 712bench the authors ran into an unexpected memory exhaustion at the 780759th allocation in vanilla Pebbles. Running the bench with d-Pebbles allowed us to step into the debug shell at failure and use the ’memstat’ command in d-Mode to find an over-population of mem region t objects, exactly 780759 of them to be precise, hinting a memory leak in remove pages. \\

\subsection{Execution Overhead with contention}
Writing VM logs by multiple threads requires synchronization. mutex712bench creates C child processes and each process performs N memory allocations and N deallocations. As each child spawned child process does memory allocations, they all contend for a slot on the VM log list, with a higher C implying higher contention. We measure the time to perform these C*N allocation in d-Pebbles with VM logging and compare it with vanilla Pebbles using getticks() to measure elapsed time. Fig 2b. shows the relative overhead of d-Pebbles by varying C and keeping N fixed at 100. We observe that the overhead due to locking is less than 4\%.

\includegraphics[scale=0.3]{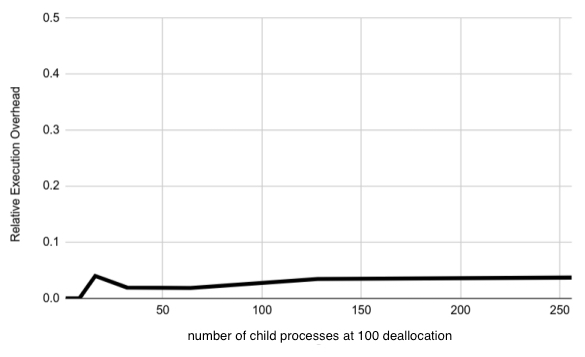}
\begin{center}
    Fig 2b. Relative execution overhead with contention of d-Pebbles for mutex712bench.
\end{center}

\hfill\\
\section{Conclusion and Future Work}
By designing, implementing and evaluating d-Pebbles, we conclude that a kernel with ”debuggability” as a design principle can indeed be constructed and it is effective in guiding its own debugging. We demonstrate that making use of our recommendations can improve the debuggability of a kernel without severe overhead. We expect the recording overhead and interrupt latency related to our record-and-replay to have high cost, but believe that exploiting hardware support would help optimize away the runtime overheads. We would also like to explore the exciting nature of record-and-replay in a multiprocessor setting as the degree of nondeterminism increases sharply when compared to a uniprocessor. Trusting a kernel to not corrupt the reserved memory area is a general concern whilst debugging without external support. We plan to look into cryptographic hashing (signing) to check integrity, and also explore hardware extensions for enclaves such as Intel’s SGX to guarantee integrity. As an end result, we would like to obviate the need for external debugging tools for kernel development and enable the kernel to be self sufficient in this regard.

\section{Acknowledgements}

We would like to thank Phil Gibbons and Andrew Chung for constant guidance right from the ideation phase for this endeavour. We were able to convert an initial unpolished idea that was something along the lines of ”the kernel debugging itself” to a full fledged study of how a kernel’s design affect its debuggability thanks to their guidance. The research papers that we studied as a part of the 15-712 course has enabled us to conduct research with a critical eye for detail.

\bibliographystyle{ieeetr}
\bibliography{bibliography}

\clearpage

\end{document}